\documentclass{Interspeech}

\interspeechcameraready

\title{Improved Intelligibility of Dysarthric Speech using Conditional Flow Matching}

\author[affiliation={1}]{Shoutrik}{Das}
\author[affiliation={1}]{Nishant}{Singh}
\author[affiliation={1}]{Arjun}{Gangwar}
\author[affiliation={1}]{S}{Umesh}

\affiliation{Spring Lab, Department of Electrical Engineering}{Indian Institute of Technology Madras}{India}
\email{dasshoutrik@gmail.com, snishant.work@gmail.com, arjungangwar@gmail.com, umeshs@ee.iitm.ac.in}

\usepackage{comment}
\usepackage{amsmath, amssymb}
\usepackage{graphicx}
\usepackage{float}
\usepackage{booktabs}

\begin{document}

\maketitle

\begin{abstract}
    
Dysarthria is a neurological disorder that significantly impairs speech intelligibility, often rendering affected individuals unable to communicate effectively. This necessitates the development of robust dysarthric-to-regular speech conversion techniques. In this work, we investigate the utility and limitations of self-supervised learning (SSL) features and their quantized representations as an alternative to mel-spectrograms for speech generation. Additionally, we explore methods to mitigate speaker variability by generating clean speech in a single-speaker voice using features extracted from WavLM. To this end, we propose a fully non-autoregressive approach that leverages Conditional Flow Matching (CFM) with Diffusion Transformers to learn a direct mapping from dysarthric to clean speech. Our findings highlight the effectiveness of discrete acoustic units in improving intelligibility while achieving faster convergence compared to traditional mel-spectrogram-based approaches.
\end{abstract}

\section{Introduction}

Effective communication is fundamental to human interaction, yet individuals with dysarthria—a motor speech disorder—encounter profound challenges in expressing themselves clearly. Dysarthria arises from neurological conditions that impair the control and coordination of the muscles involved in speech production. As a result, affected individuals may experience slurred articulation, imprecise pronunciation, reduced speech volume, and altered speech rhythm, making verbal communication difficult for both the speaker and the listener.

Dysarthria is not a singular condition but rather a broad classification encompassing multiple types, each differing in severity and underlying cause. It commonly results from neurological disorders such as Parkinson’s disease, amyotrophic lateral sclerosis (ALS), stroke, traumatic brain injury, and other motor neuron diseases. These conditions disrupt the intricate neural pathways responsible for coordinating the movement of the tongue, lips, vocal cords, and respiratory muscles, leading to varying degrees of speech impairment.

Dysarthria severity varies from mild articulation imprecision to complete speech loss, depending on the disorder, disease stage, and affected speech regions. Parkinsonian dysarthria often causes monotony, breathiness, and low volume (80–100\% of cases), while ALS-related dysarthria can lead to total speech loss (up to 30\%). Stroke-induced dysarthria varies based on brain damage, affecting speech clarity to different extents.

Due to the varying degrees of speech impairment in dysarthria, the need for converting impaired speech into intelligible, natural-sounding speech is of paramount importance. In this work, we propose a method for Dysarthric speech conversion(DSC) without requiring a paired corpus. Our approach leverages the associated text and a state-of-the-art text-to-speech (TTS) model \cite{f5} to synthesize regular speech for a given dysarthric utterance.

Furthermore, we introduce a conditional-flow-matching\cite{flow} based model to learn the mapping between the units of dysarthric speech and the corresponding regular speech Mel spectrogram. The generated spectrogram is then processed by vocoders such as BigVGAN \cite{bigvgan} to synthesize high-quality speech. Additionally, we demonstrate that using quantized units as an intermediate representation significantly outperforms direct Mel spectrogram conversion. This improvement can be attributed to the high variability in Mel spectrograms, which makes direct transformation less effective for producing natural and intelligible speech.

\section{Related work}

Converting Dysarthric speech to regular speech presents two primary challenges: the absence of parallel dysarthric-clean speech corpus and the complexity of learning a robust transformation function.

The first major hurdle in developing dysarthric-to-regular speech conversion (DSC) systems is the inherent inavailability of a parallel dysarthric-clean speech corpus for supervised learning. Since a single speaker cannot simultaneously produce both dysarthric and regular speech, the construction of a parallel corpus is fundamentally constrained. To overcome this, researchers have explored various strategies. Some studies propose modifying the speech rate and rhythm of typical speech to simulate dysarthric characteristics \cite{vachhani}, while others employ voice conversion (VC) techniques that adjust the prosodic and formant structures of dysarthric speech to align more closely with those of non-dysarthric speakers \cite{hosom}. More recently, deep learning-based methods, particularly adversarial training frameworks, have been leveraged to tackle this challenge \cite{gan_for_dys}.  Notable advancements in this direction include Dysarthric Voice Conversion (DVC) systems \cite{dvc, dvc3}, which integrate TTS technology with the StarGAN-VC architecture to enhance speech transformation quality.

Even if a parallel corpus were readily available, the next challenge lies in learning a robust function that maps dysarthric speech to regular speech. This function must effectively transform dysarthric speech while preserving its phonetic content and semantics, eliminating only the disfluencies introduced by the disorder. Generative Adversarial Networks (GANs) \cite{gan0, gan1}, have emerged as a leading approach for this task, particularly in speech enhancement \cite{gan_for_se1, gan_for_se2, gan_for_se3, segan}. By leveraging GAN-based frameworks, researchers have been able to improve the intelligibility of dysarthric speech, making it more suitable for both human listeners and automatic speech recognition (ASR) systems. The work that closely aligns with our approach is the paper called Unit-DSR \cite{unitDSR}, which utilizes a speech unit normalizer and a Unit HiFi-GAN vocoder.

\section{Method}

In this section, we outline the methodology employed in our work. We begin by describing our approach for generating regular speech from dysarthric speech. Next, we introduce the quantization technique used to obtain discrete acoustic units and discuss their relationship to various speech characteristics. Following this, we present our proposed model for learning the transformation from dysarthric to regular speech. We then detail our pretraining methodology, which serves as a foundation for effective weight initialization. Finally, we conclude with our fine-tuning strategy, which further optimizes the system for dysarthric to clean speech conversion.

\subsection{Data Generation}

For data generation, we employ the F5-TTS \cite{f5} architecture, a Diffusion Transformer-based model trained using the flow-matching technique. Unlike traditional TTS models, F5-TTS eliminates the need for a complex alignment mechanism, such as a duration predictor. Our choice of this architecture is driven by its ability to synthesize speech with any desired speaker characteristics using only a few seconds of reference audio. Furthermore, it incorporates the Sway Sampling strategy, achieving an inference real-time factor (RTF) of 0.15, making it highly efficient.

To generate clean speech, the corresponding text for each dysarthric audio sample from the Speech Accessibility Project dataset[SAPC] \cite{sapc} is fed into F5-TTS with a single-speaker prompt. This minimizes speaker variation at the clean speech generation stage. The following section details our choice of model for input feature generation, which further mitigates speaker variability. The underlying objective of this work is to generate intelligible speech in a single, consistent speaker voice from any dysarthric speech sample, regardless of the speaker's identity or severity of dysarthria. To mitigate speaker variability, we use single-speaker synthesized speech, ensuring better generalization in DSC models.

\subsection{Acoustic Discrete Units}
Self-supervised speech models (S3Ms) excel at capturing various linguistic features, including phonetics, semantics, and morphology. A previous study \cite{units2} has shown that in the hidden representations of these models, phonetically similar words are grouped more closely than semantically similar words. Additionally, models like HuBERT \cite{hubert} and WavLM \cite{wavlm} have demonstrated that phonetic and word-level information tends to cluster in the higher layers when trained on discrete hidden units \cite{karens_paper}. Furthermore, discretization has been proposed as an information bottleneck, effectively separating speaker-specific features from the linguistic content \cite{discrete1}. Another study \cite{units1} shows representing each phonetic category as a distribution over discrete units, discrete acoustic units can be characterised as sub-phonemic events, rather than high-level categories such as phonemes. 

Due to the given advantages of discrete acoustic units in this work, we leverage discrete acoustic units from WavLM to represent dysarthric speech. The key advantage of this choice is that, in contrast to HuBERT as presented in the work of Unit-DSR, WavLM is trained not only for masked unit prediction but also for speech denoising. During its training, batches are augmented with random background noise and secondary speakers, while the target units are derived from HuBERT cluster IDs. By jointly solving the tasks of masked speech denoising and prediction, the model effectively learns to distinguish the main speaker’s speech from noise and interference. This approach eliminates the need for a separate speech enhancement module, unlike techniques such as S-SEGAN \cite{s_segan}, which require additional enhancement stages before source separation.

WavLM employs a feature encoder consisting of local temporal convolutional blocks, where each output feature vector represents a 25 ms segment of audio with a stride of 20 ms. This is followed by a Transformer encoder enhanced with a gated relative position bias. For this work, we extract features from the 21st layer of WavLM-Large, for a combination of LibriSpeech \cite{libri}, GigaSpeech \cite{giga}, and VoxPopuli \cite{voxpopuli}. To ensure a balanced representation, we selected 5\% of English data from each of these datasets in equal proportions. A K-Means model with 512 clusters was then trained on these features, and the resulting cluster IDs were used as discrete representations for the dysarthric speech.

\begin{figure*}[t]
    \centering
    \hspace{-1cm} 
    \includegraphics[width=\textwidth]{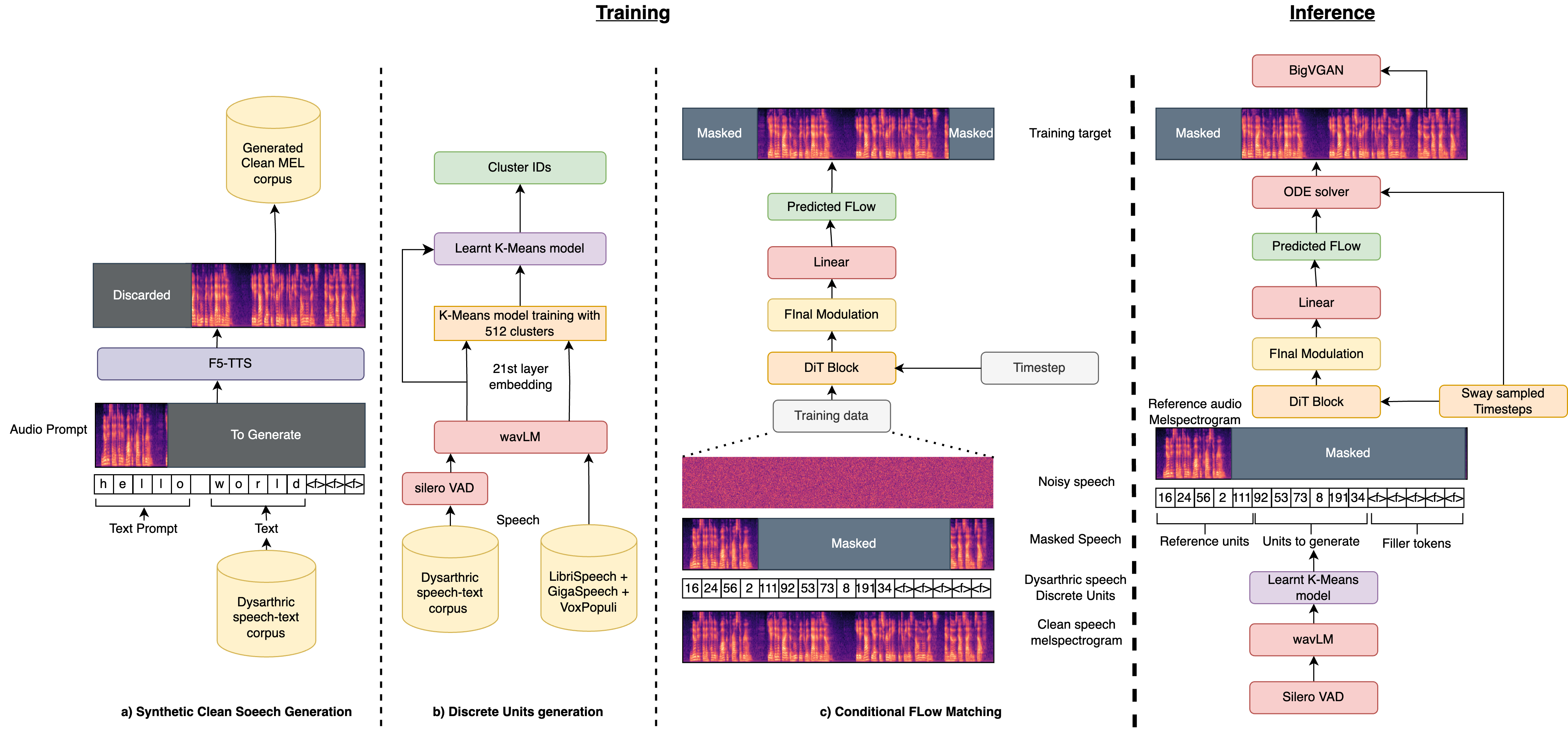}
    \caption{Diagram of the Dysarthric speech recognition system}
    \label{fig:wide_image}
\end{figure*}

\subsection{Conditional Flow Matching}

Given a flow $\varphi_t$, the probability path $p_t: [0,1] \times \mathbb{R}^d \to \mathbb{R}_{>0}$, which represents the time-dependent probability density function, can be obtained via the formula for change of variables:

\begin{equation}
    p_t(x) = p_0(\varphi_t^{-1}(x)) \det \left( \frac{\partial \varphi_t^{-1}}{\partial x} (x) \right)
\end{equation}

To sample from $p_t(x)$, one first draws $x_0$ from $p_0$ and then solves the initial value problem (IVP) for $\varphi_t(x_0)$ using the ODE formulation $d\varphi_t(x)/dt = v_t(\varphi_t(x))$ with $\varphi_0(x) = x_0$.

Let $p_t$ be a probability path with the corresponding vector field $u_t$. The vector field $v_t(x;\theta)$, parameterized by a neural network with parameters $\theta$, can be optimized using the Flow Matching (FM) objective:

\begin{equation}
\mathcal{L}_{CFM}(\theta) = \mathbb{E}_{t, p_t(x)} \lVert u_t(x) - v_t(x;\theta) \rVert ^2
\end{equation}

where $t \sim U[0,1]$ and $x \sim p_t(x)$. Despite its simplicity, directly computing this objective is infeasible due to the lack of prior knowledge of $p_t$ and $u_t$. Consequently, it is not possible to estimate the loss or its gradients directly.

To address this challenge, \cite{flow-matching} proposed constructing $p_t(x)$ using a mixture of simpler conditional paths $p_t(x | x_1)$, where the corresponding vector field $u_t(x | x_1)$ can be efficiently computed. A conditional path is defined such that: (1) $p_0(x | x_1) = p_0(x)$ and (2) $p_1(x | x_1) = \mathcal{N}(x | x_1, \sigma^2 I)$, a Gaussian distribution centered at $x_1$ with a small variance $\sigma^2$ (typically $10^{-5}$). The marginal path is calculated as $\int p_t(x | x_1) q(x_1) dx_1$, which closely approximates $q(x_1)$ at $t = 1$.

Based on this, the Conditional Flow Matching (CFM) objective was introduced:

\begin{equation}
    \mathcal{L}_{CFM}(\theta) = \mathbb{E}_{t, q(x_1), p_t(x|x_1)} \lVert u_t(x | x_1) - v_t(x;\theta) \rVert ^2
\end{equation}

It has been established that FM and CFM yield identical gradients with respect to $\theta$. More importantly, CFM enables efficient sample generation from $p_t(x | x_1)$ and unbiased gradient estimation.

A critical aspect of CFM is the choice of the conditional flow. Since a flow defines trajectories that transition from $p_0$ to $p_1$, simpler trajectories (e.g., straight-line paths) can enhance learning efficiency and improve IVP-solving accuracy. \cite{flow-matching} proposed an optimal transport (OT) path, the time-dependent conditional probability path and corresponding vector field is given by:

\begin{equation}
    p_t(x | x_1) = \mathcal{N}(x | t x_1, (1 - (1 - \sigma_{\min}) t)^2 I),
\end{equation}

\begin{equation}
    u_t(x | x_1) = \frac{x_1 - (1 - \sigma_{\min})x}{1 - (1 - \sigma_{\min})t}.
\end{equation}

This formulation ensures a simple, linear motion of points, allowing for more efficient training. We adopt this approach in this paper. 

Given a dataset $(x,y)$ where $x \in \mathbb{R}$ denotes the audio sample and $y \in \mathbb{Z}^+$ denotes the discrete acoustic units, we extract log mel-spectrogram, $x' \in \mathbb{R}^{80}$ from the audio sample.  We add filler tokens to $y'$ to match the length of $x'$. The units are then converted into embeddings, $y' \in \mathbb{R}^{512}$ using a embedding look-up table. Let $m$ be a binary temporal mask which is of the same length as $x'$, then $x'_{mis} = m \odot x'$ and $x'_{ctx}=(1-m) \odot x'$.  The conditional flow matching model is learning $p(x'_{mis} \mid y',x'_{ctx})$.

We train the model in two stages : Pre-training and Fine-tuning, discussed in following sections. 

\subsection{Training}
The input speech is first processed using WavLM, from which we extract features from the 21st layer. These features are then quantized by mapping each feature vector to the nearest centroid in a pretrained K-Means model, using Euclidean distance as the criterion. Each centroid is represented by an integer, referred to as a units in this work. As different speakers exhibit varying speech rates, the degree of repetition for each units differs accordingly. To normalize this variation, we apply a discrete accoustic units collapsing strategy, wherein consecutive occurrences of the same units are merged into a single instance. These collapsed discrete representations, along with the masked mel-spectrogram of the clean speech and the noisy mel-spectrogram, serve as inputs to the Diffusion Transformer module, which is trained using the Conditional Flow Matching (CFM) algorithm (detailed in Section \textbf{3.3}).

\subsubsection{Pretraining}
To leverage the abundance of high-quality, professionally recorded speech data available online, we pretrain the model on the 960-hour LibriSpeech dataset. This stage utilizes both the extracted units and the corresponding mel-spectrogram representations derived from the same speech signal. The motivation behind this approach is the significant disparity in the availability of clean speech data compared to dysarthric speech. The pretrained weights obtained from this phase serve as the initialization for the fine-tuning stage.

\subsubsection{Finetuning}
We finetuned the pretrained model on the Speech Accessibility Project dataset \cite{sapc}. The fine-tuning process follows the same framework as pretraining, with the key distinction that the input acoustic units are derived from dysarthric speech, while the target mel-spectrogram corresponds to the artificially generated single-speaker clean speech. Additionally, during fine-tuning, we apply voice activity detection (VAD) to remove unnecessary silence segments from the dysarthric speech. Specifically, we focus on eliminating leading and trailing silences, setting a higher speech-probability threshold for these regions compared to intra-speech silences.

\subsection{Inference} 
During inference, the dysarthric speech is processed using Silero VAD, followed by WavLM feature extraction and unit generation via a pretrained K-Means model. These units, along with a reference clean mel spectrogram, are fed into the Conditional Flow Matching (CFM) module to generate the target mel spectrogram using an infilling-based approach. We follow F5-TTS's inference strategy, employing an ODE solver and Sway Sampling for efficient speech synthesis.

\section{Training setup}
All speech signals were resampled to 16 kHz during both feature extraction with WavLM and the training process. For the target representation, mel spectrograms were generated using a window length of 40 ms, hop length of 10 ms, and an FFT size of 1024 points, with 80 mel channels to maintain compatibility with our vocoder, BigVGAN, trained on 100 hours of LibriSpeech Dataset. In this study, we pre-trained and fine-tuned two models, referred to as CFM-base and CFM-small. The CFM-base model employs a Diffusion Transformer architecture consisting of 18 layers, 12 attention heads, and a model dimension of 768. In contrast, the CFM-small model comprises 9 layers, 6 attention heads, and a model dimension of 512. The CFM-base model was pre-trained for 600,000 updates, with each update corresponding to approximately 42 minutes of audio, while the CFM-small model was pre-trained for 400,000 updates, where each update used roughly 20 minutes of audio.

We optimized both models using the AdamW optimizer \cite{adamw}, setting a peak learning rate of 7.5e-5 during pre-training and 1e-5 during fine-tuning. The learning rate was scheduled with a warmup phase of 20,000 steps for pre-training and 10,000 steps for fine-tuning. Fine-tuning continued for an additional 400,000 updates on the base model and 300,000 updates on the small model. For computational resources, the CFM-base model was trained on four NVIDIA RTX 6000 GPUs, whereas the CFM-small model was trained on two NVIDIA RTX 6000 GPUs.

\section{Experimental Results}
We evaluate our model on a subset of the Speech Accessibility Project dataset reserved for testing, which includes speech from both seen and unseen speakers. As baselines, we use wav2vec2-base-960h \cite{wav2vec2} and its finetuned counterpart trained on the same 
SAPC dataset. We assess transcription quality by evaluating wav2vec2 (without finetuning) on speech generated by our three models. Additionally, we conduct a Mean Opinion Score (MOS) evaluation, where 23 self-reported English speakers rate 10 randomly sampled test utterances on a scale of 1 to 5, with 1 indicating poor quality and 5 representing higher intelligibility.

\begin{table}[H]
\centering
\begin{tabular}{lll}
\toprule
Method           & WER  & MOS \\  
\midrule \midrule
CFM-base-units    & $31.30$  & $\textbf{3.9}$ \\  
CFM-small-units   & $34.69$  & $3.6$ \\  
CFM-base-MEL      & $84.07$  & $1.0$ \\  
wav2vec2             & $39.02$  & - \\  
wav2vec2-finetuned         & $\textbf{24.82}$  & - \\  
\bottomrule
\end{tabular}
\caption{Average WER (\%) and MOS comparison for SAPC test set.}
\label{tab:results1}
\end{table}

\begin{table}[H]
\centering
\begin{tabular}{llll}
\toprule
Speaker & WER & MOS & Confidence Interval\\  
\midrule \midrule
Seen & $18.2$ & $3.8$ & $3.65-4.05$\\  
Unseen & $44.40$ & $3.7$ & $3.47-3.87$\\  
Unseen finetuned & $30.90$ & $1.0$ & $1.00-1.00$\\  
\bottomrule
\end{tabular}
\caption{WER (\%) and MOS comparison between seen, unseen, and finetuned unseen speakers.}
\label{tab:results2}
\end{table}

To further examine the impact of input representations, we trained the base model using mel-spectrograms of dysarthric speech instead of discrete acoustic units. While the model trained with discrete units achieved intelligible speech after 1M updates (combined pretraining and finetuning), the mel-spectrogram-based model failed to do so even after 2M updates. This discrepancy can be attributed to differences in temporal alignment: dysarthric mel-spectrograms are typically longer than their clean counterparts, necessitating the use of a learned padding embedding for target alignment. In contrast, discrete units, after collapsing consecutive repetitions, operate at a lower temporal resolution, similar to text-to-speech (TTS) systems such as F5-TTS, where unit sequences are padded with learned embeddings.

We hypothesize that when the input is shorter than the target, the model can begin learning input-output alignment at lower layers of the Diffusion Transformer (DiT). However, with mel-spectrograms, the compression is deferred to higher layers, slowing convergence. Additionally, units introduce a speaker-invariant bottleneck, simplifying the mapping to clean speech, whereas mel-spectrograms inherently encode speaker-specific attributes, requiring an implicit speaker normalization step. This added complexity increases training difficulty and delays convergence. The observed Word Error Rate (WER) differences further support this, as units enable more precise phonetic reconstruction, improving transcription accuracy.

We also examined the impact of speaker familiarity on performance. Our models performed significantly better when tested on speakers included in the training set compared to unseen speakers, as reflected in the MOS results. However, we found that this limitation could be overcome by fine-tuning the model with just 1 hour of speech from the target speaker. This indicates that the model retains a strong ability to adapt with minimal additional data, which is particularly promising for real-world deployment scenarios where only a small amount of clean speech from a new speaker may be available.

\section{Conclusion}
We propose the use of acoustic discrete units as an alternative to mel-spectrograms for dysarthric-to-clean speech generation. Our experiments show that acoustic discrete units not only achieve significantly faster convergence than mel-spectrograms but also produce speech with higher intelligibility. Additionally, we demonstrate that Conditional Flow Matching (CFM) serves as a viable alternative to Generative Adversarial Networks (GANs) and traditional signal processing methods for dysarthric speech conversion. This study represents a preliminary step toward leveraging CFM and discrete units for dysarthric speech restoration. In future work, we plan to conduct a more comprehensive evaluation across a wider range of dysarthric conditions and speakers, as well as extend our approach to non-English languages.

\bibliographystyle{IEEEtran}
\bibliography{mybib}

\end{document}